\documentclass[%
 aip,
 jmp,%
 amsmath,amssymb,
 reprint,%
]{revtex4-1}

\usepackage{graphicx}
\usepackage{dcolumn}
\usepackage{bm}
\usepackage[sort&compress]{natbib}
\usepackage{xspace}
\usepackage{amsmath}
\usepackage{amssymb}
\usepackage{epsfig}

\begin{document}

\title{Integrated tuning fork nanocavity optomechanical transducers with high $\bm{f_{M}Q_{M}}$ product and stress-engineered frequency tuning}

\author{R. Zhang}
\affiliation{Department of Mechanical Engineering, Worcester Polytechnic Institute, Worcester, MA 01609, USA}
\author{C. Ti}
\affiliation{Department of Mechanical Engineering, Worcester Polytechnic Institute, Worcester, MA 01609, USA}
\author{M. I. Davan\c co}
\affiliation{Center for Nanoscale Science and Technology, National
Institute of Standards and Technology, Gaithersburg, MD 20899,
USA}
\author{Y. Ren}
\affiliation{Department of Mechanical Engineering, Worcester Polytechnic Institute, Worcester, MA 01609, USA}
\author{V. Aksyuk}
\affiliation{Center for Nanoscale Science and Technology, National
Institute of Standards and Technology, Gaithersburg, MD 20899, USA}
\author{Y. Liu} \email{yliu11@wpi.edu}
\affiliation{Department of Mechanical Engineering, Worcester Polytechnic Institute, Worcester, MA 01609, USA}
\affiliation{Center for Nanoscale Science and Technology,
National Institute of Standards and Technology, Gaithersburg, MD
20899, USA}
\author{K. Srinivasan}
\affiliation{Center for Nanoscale Science and Technology, National
Institute of Standards and Technology, Gaithersburg, MD 20899, USA}

\date{\today}

\begin{abstract}
Cavity optomechanical systems are being widely developed for precision force and displacement measurements. For nanomechanical transducers, there is usually a trade-off between the frequency ($f_{M}$) and quality factor ($Q_{M}$), which limits temporal resolution and sensitivity. Here, we present a monolithic cavity optomechanical transducer supporting both high $f_{M}$ and high $Q_{M}$. By replacing the common doubly-clamped, Si$_3$N$_4$ nanobeam with a tuning fork geometry, we demonstrate devices with the fundamental $f_{M}\approx29$~MHz and $Q_{M}\approx2.2$$\times10^5$, corresponding to an $f_{M}Q_{M}$ product of 6.35$\times10^{12}$~Hz, comparable to the highest values previously demonstrated for room temperature operation. This high $f_{M}Q_{M}$ product is partly achieved by engineering the stress of the tuning fork to be 3 times the residual film stress through clamp design, which results in an increase of $f_{M}$ up to 1.5 times. Simulations reveal that the tuning fork design simultaneously reduces the clamping, thermoelastic dissipation, and intrinsic material damping contributions to mechanical loss. This work may find application when both high temporal and force resolution are important, such as in compact sensors for atomic force microscopy.
\end{abstract}
\maketitle

Cavity optomechanical systems are being developed for many applications in precision force and displacement measurements~\cite{ref:Kippenerbg_Vahala_Science,ref:Metcalfe_optomechanics}. Monolithic systems in which nanomechanical transducers are combined with integrated optical readout have been developed in geometries where optical resonances and mechanical modes are co-located within the same physical structure~\cite{ref:Antoni_deformable_slab,ref:eichenfield1,ref:Forstner_Bowen_magnetometer,ref:Kim_Carmon,ref:Wu_Barclay_dissipative}, and in systems for which optical and mechanical modes are supported by different physical structures and are near-field-coupled~\cite{ref:Gavartin_Kippenberg_force_transducer,ref:Liu_Srinivasan_wide_stiffness1}. Such near-field coupling enables the mechanical resonator size to be scaled down to the nanoscale while maintaining high displacement sensitivity~\cite{ref:Gavartin_Kippenberg_force_transducer,ref:Liu_Srinivasan_wide_stiffness1}in contrast to far-field optical readout, where diffraction effects limit the mechanical resonator size that can be sensitively detected~\cite{ref:Kouh_Ekinci_interferometer}. For a given desired mechanical stiffness (determined by the force sensing application), a nanoscale cantilever can have much higher resonant frequency $f_{M}$ (and therefore transduction bandwidth/temporal resolution) than a microscale counterpart, due to its smaller effective motional mass ($m$). Such a high frequency mechanical resonator would ideally exhibit a high mechanical quality factor ($Q_{M}$), as the force sensitivity scales as $1/(f_{M}^{1/2}Q_{M}^{1/2})$ (Ref.~\onlinecite{ref:yasumura2000quality}). However, there is usually a tradeoff between $f_{M}$ and $Q_{M}$ because shrinking down the resonator size comes at the expense of a reduction in $Q_{M}$, due to increased clamping losses~\cite{ref:yasumura2000quality,ref:verbridge_nano_letters}. Here, we demonstrate an integrated silicon nitride cavity optomechanical system where high $f_{M}$ and $Q_{M}$ are simultaneously achieved. Using microdisk optical resonators with intrinsic optical quality factor $Q_{o}>6{\times}10^5$ for readout, we develop a doubly-clamped tuning fork geometry as the mechanical resonator, where we take advantage of elastic wave interference to limit the mechanical loss, while retaining the high tensile stress characteristic of stoichiometric Si$_{3}$N$_{4}$. We investigate different clamp designs that can increase the tensile stress by up to 2.9 times and hence tune $f_{M}$ on chip, as well as surface treatment to improve $Q_{M}$. Devices with $f_{M}\approx29$~MHz and $Q_{M}\approx2.2{\times}10^5$ are presented, corresponding to an $f_{M}Q_{M}$ product of $6.35{\times}10^{12}$~Hz.

\begin{figure}[h]
\centerline{\includegraphics[width=\linewidth]{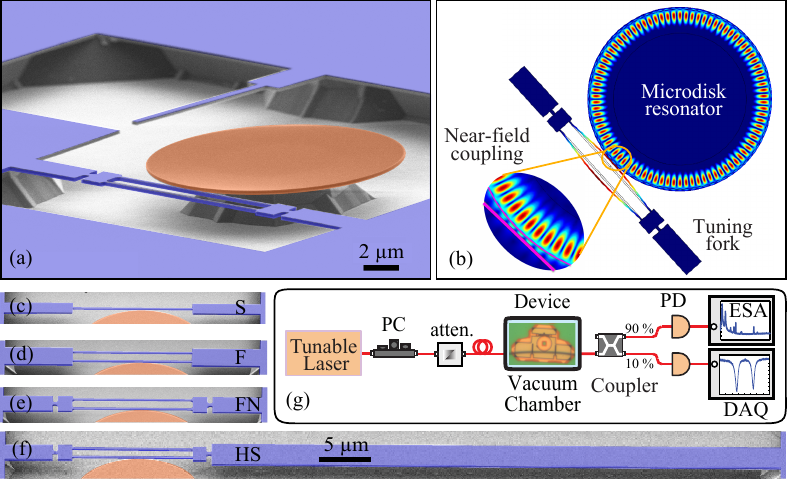}}
\caption{(a) False-colored scanning electron micrograph (SEM) of a fabricated device. The orange region indicates the Si$_3$N$_4$ microdisk optical resonator while the blue region indicates the mechanical resonator and supporting structures. (b) Working principle of the device. The colored tuning fork shows the calculated mode shape of the first-order, in-plane, out-of-phase mechanical mode. The colored disk shows the TE$_{1,43}$ whispering gallery optical mode. (c)-(f) SEM images of four different mechanical resonator designs: (c) single beam (S), (d) tuning fork (F), (e) tuning fork with neck (FN), (f) high-stress tuning fork (HS). (g) Schematic of the characterization system. PC, polarization controller; PD, photodiode; DAQ, data acquisition; ESA, electrical spectrum analyzer.}
\label{fig:fig1}
\end{figure}

Doubly-clamped silicon nitride nanobeams have been extensively explored as nanomechanical resonators, with the high residual tensile stress ($\approx$1~GPa) produced by low-pressure chemical vapor deposition (LPCVD) of stoichiometric silicon nitride (Si$_{3}$N$_{4}$) on silicon enabling $Q_{M}>10^6$ to be achieved for MHz frequency modes~\cite{ref:Verbridge}. In the context of cavity optomechanics, such high-$Q_{M}$ structures have been evanescently coupled to silica microdisk resonators~\cite{ref:Gavartin_Kippenberg_force_transducer} and incorporated within silicon nitride nanophotonic circuits~\cite{ref:Fong_Tang}. Here, our goal is to develop a Si$_{3}$N$_{4}$ nanobeam-microdisk optomechanical system in which the mechanical frequency $f_{M}$ is increased to the tens of MHz range while maintaining high $Q_{M}$. To do so, we replace the commonly used single beam resonator with a tuning fork mechanical resonator, as shown in Fig.~\ref{fig:fig1}(a).

For nanomechanical resonators operating in vacuum, there are several sources of mechanical energy dissipation: clamping losses ($Q_{clamp}$), thermoelastic dissipation (TED) ($Q_{TED}$), and material losses ($Q_{mat}$). The total mechanical quality factor $Q_{M}$ can then be written as $1/Q_{M}=1/Q_{clamp}+1/Q_{TED}+1/Q_{mat}+1/Q_{other}$ (Ref.~\onlinecite{ref:yasumura2000quality}). Clamping losses occur when the elastic energy radiates into its support structures~\cite{ref:yasumura2000quality,ref:Hao_Ayazi}. TED is the energy dissipation due to strain-induced heating and the resulting temperature gradients~\cite{ref:Duwel,ref:lifshitz_roukes_TED}, and is expected to dominate Akhesier damping for devices with our geometry/aspect ratios (so that the latter is not considered further in our discussion)~\cite{ref:unterreithmeier2010damping}. In addition, losses caused by localized defect states both on the surface and in the volume of the material cannot be neglected~\cite{ref:unterreithmeier2010damping}. Compared with the modes of single cantilever devices~\cite{ref:Gavartin_Kippenberg_force_transducer,ref:Liu_Srinivasan_wide_stiffness1,ref:Fong_Tang} the in-plane, out-of-phase mechanical mode of the tuning fork structure (mode shape shown in Fig.~\ref{fig:fig1}(b)) enables the total force and moment at the outer end of the clamps to be zero. Therefore, the motion of the beams is effectively decoupled from the clamps, resulting in an expected lower clamping losses. This effect, which has been used in a variety of mechanical structures ranging from clocks and musical instruments (the tuning fork was invented in 1711~\cite{ref:feldmann1997history}) to atomic force microscope sensors~\cite{ref:Giessibl_RMP}, can also be understood as destructive interference of elastic waves, and similar ideas have been implemented in double disk cavity optomechanical structures~\cite{ref:Zhang_Lipson_anchor_loss}. In contrast to those structures, the doubly clamped geometry we adopt (Fig.~\ref{fig:fig1}(a)-(b)) maintains the high residual tensile stress of the Si$_3$N$_4$ film, which is important for maintaining high $f_M$ and $Q_M$~\cite{ref:verbridge_nano_letters}.

To verify these benefits of the design, we fabricated cavity optomechanical transducers with four types of mechanical resonators: single beams (S), turning forks (F), and tuning forks with necks (FN), as shown in Fig.~\ref{fig:fig1}(c)-(e). The tuning fork with and without neck geometries are nearly identical, with only the specifics of the clamping geometry slightly modified by inserting a neck region. Each mechanical resonator was coupled to a microdisk optical resonator that is 15~$\mu$m in diameter, and the separation between the optical and mechanical resonators is 150~nm. The devices were fabricated in a 250~nm stoichiometric silicon nitride film with an internal tensile stress of $\approx$~1.1~GPa, using electron-beam lithography and dry etching processes similar to our previous work on high-$Q$ Si$_3$N$_4$ microdisk optomechanical resonators~\cite{ref:Liu_yuxiang_wlc}. The mechanical resonator beams are 150~nm wide, and the beam lengths are varied between 12~$\mu$m and 40~$\mu$m to investigate the dependence of device performance on the beam length. Stress tuning is an effective way to increase both $f_{M}$ and $Q_{M}$, and stress tuning of chip-based devices was previously realized by substrate bending~\cite{ref:verbridge_nano_letters}. In the tuning fork design, a stress level within the mechanical resonator above that of the deposited film was achieved by increasing the clamp length on one side of the fork, for example, by 100~$\mu$m as shown in Fig.~\ref{fig:fig1}(f). Due to the initial unbalanced tensile forces in the clamp and beam (details available in supplementary materials~\cite{ref:tuning_fork_note}), the stress is redistributed after undercut, such that it decreases in the wide suspended clamp and increases by up to 3 times~\cite{ref:tuning_fork_note} in the attached narrow beams of the tuning fork. To our knowledge, this is the first time stress tuning has been achieved by a design-enabled, on-chip approach. This stress tuning method provides more flexibility and better control of engineering $f_{M}$ of nanomechanical resonators.

\begin{figure}[b!]
\centerline{\includegraphics[width=\linewidth]{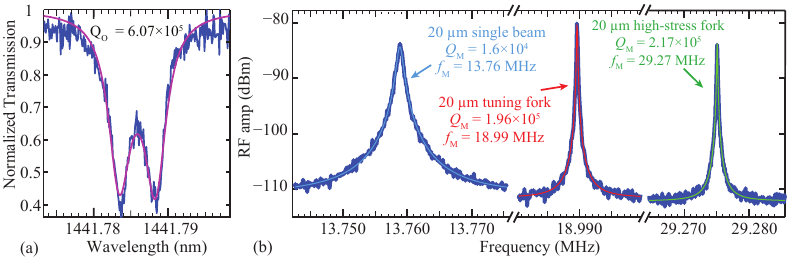}}
\caption{(a) Typical optical spectrum of microdisk resonator. The intrinsic optical Q factor is $6.07{\times}10^5~\pm~6{\times}10^3$. The 95~$\%$ confidence interval range for $Q_{o}$ is determined by a nonlinear least squares fit to the data. (b) Mechanical spectra of a single beam, tuning fork, and high-stress tuning fork structures with 20~$\mu$m beam length. The 95~$\%$ confidence intervals for $Q_{M}$ from the curve fitting are typically $\pm3~\%$, and are summarized in Fig.~\ref{fig:fig3}(a)-(b).}
\label{fig:fig2}
\end{figure}

For each mechanical resonator geometry shown in Fig.~\ref{fig:fig1}(c)-(f), devices with three different beam lengths (12~$\mu$m, 20~$\mu$m, and 40~$\mu$m) were fabricated and characterized in vacuum (0.13~Pa) in order to evaluate both their $f_{M}$ and $Q_{M}$. The characterization setup is shown in Fig.~\ref{fig:fig1}(g), with more details available in the supplementary material~\cite{ref:tuning_fork_note}. The typical optical quality factor $Q_{o}$ is $>10^5$, as shown in Fig.~\ref{fig:fig2}(a). The optomechanical coupling coefficient ($g_{OM}/2\pi$) of our device is calculated to be 140 MHz/nm by perturbation theory~\cite{ref:eichenfield12} using finite element method (FEM) determined mode solutions (Fig.~\ref{fig:fig1}(b)), with details provided in the supplementary material~\cite{ref:tuning_fork_note}. Figure~\ref{fig:fig2}(b) shows the measured mechanical spectra of devices with the same beam length (20~$\mu$m) but different geometries. We find that the tuning fork with and without neck increase $Q_{M}$ by 12 times and $f_{M}$ by 1.4 times the values of a single beam structure, respectively. The high-stress tuning fork device with a longer clamp produces an even more pronounced increase in $Q_{M}$ and $f_{M}$, by 13.5 times and 2.1 times the values of a single beam structure, respectively. The stress-tuning induced $f_{M}$ increase is 1.5 times compared with regular tuning forks. We note that this device exhibited the highest $f_{M}Q_{M}$ product ($6.35{\times}10^{12}$~Hz) for any of the structures we have investigated.

The thermomechanical force noise is also decreased using the tuning fork geometry. Assuming the force noise is frequency independent, its spectral density can be estimated by $S_{F}=4kk_{B}T/(2\pi f_{M}Q_{M})$ (Ref.~\onlinecite{ref:yasumura2000quality}), where $k$ is the cantilever effective spring constant, $k_{B}$ is the Boltzmann constant, and $T$ is the temperature. For 20~$\mu$m long single beam, fork, and high-stress fork devices, the FEM calculated $k$ are 13.23~N/m, 26.27~N/m, and 60.58~N/m, respectively. Using the calculated $k$ values and the measured $f_{M}$ and $Q_{M}$ values, the corresponding $S_{F}^{1/2}$ are 0.40~fN/Hz$^{1/2}$, 0.14~fN/Hz$^{1/2}$ and 0.16~fN/Hz$^{1/2}$, respectively. These results show that, although $k$ is increased by the tuning fork design, the high $f_{M}Q_{M}$ product still reduces the force noise by a factor of $\approx$2.9 compared with the single beam devices. We also note that the mechanical damping coefficient $\gamma = k/(2\pi mf_{M}Q_{M})$ and the corresponding thermodynamic force noise $S_{F}=4k_{B}Tm\gamma $ do not further decrease  by the high stress tuning fork, which introduces an additional stress increase of 2.3 times compared with regular fork~\cite{ref:tuning_fork_note}, as the increase in $f_{M}Q_{M}$ is offset by the increase in $k$.

\begin{figure}[t]
\centerline{\includegraphics[width=\linewidth]{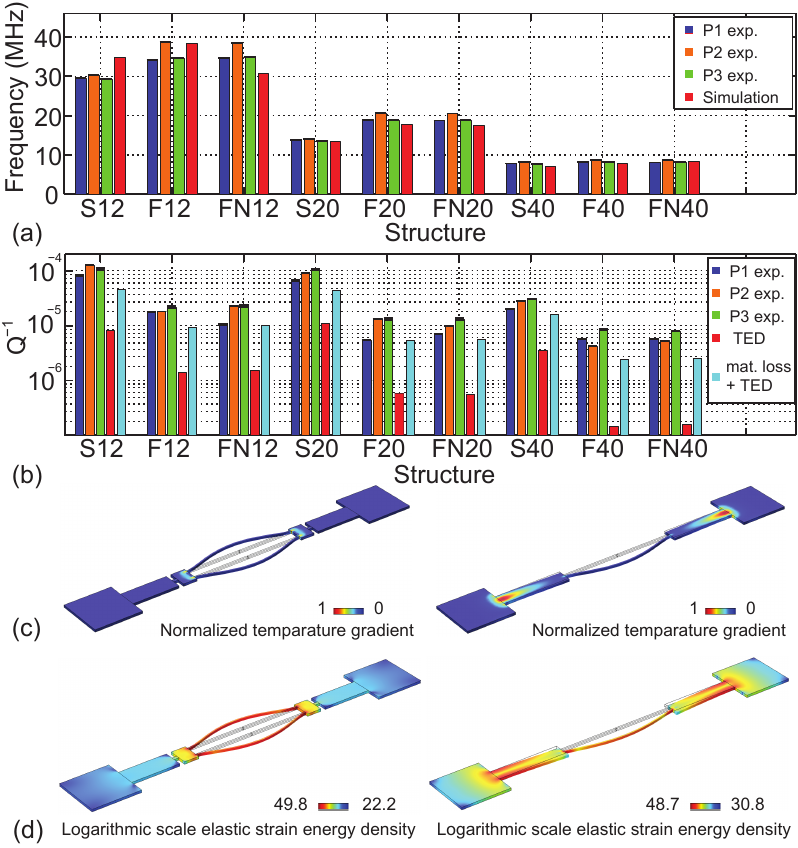}}
\caption{(a) FEM calculated and experimentally measured frequencies of the mechanical resonators, where the labels P1, P2, and P3 indicate experimental results from different batches of devices. The horizontal axis labels S, F, and FN stand for single beam, tuning fork, and tuning fork with neck, respectively, while 12, 20, and 40 specify the beam length in micrometers. (b) Experimentally measured $Q_{M}^{-1}$, FEM calculated $Q_{TED}^{-1}$, and FEM calculated $Q_{TED}^{-1}+Q_{mat}^{-1}$. The 95~$\%$ confidence interval ranges for $f_{M}$ and $Q_{M}$, determined by a nonlinear least squares fit to the data, are denoted by the solid black regions above each bar in the graphs. (c) FEM calculated normalized temperature gradient for tuning fork and single beam structures during vibration. (d) FEM calculated elastic strain energy density on a logarithmic scale for tuning fork and single beam structures during vibration.}
\label{fig:fig3}
\end{figure}

The full parametric study of single beam, tuning fork, and tuning fork with neck structures is shown in Fig.~\ref{fig:fig3}, where the labels P1, P2, and P3 indicate experimental results from different batches of devices. As shown in Fig.~\ref{fig:fig3}(a), the experimentally measured $f_{M}$ values match the FEM simulation results, indicating the measured modes were the in-plane, out-of-phase mechanical modes. Figure~\ref{fig:fig3}(a) further shows that the measured $f_{M}$ is higher in the tuning fork device architectures compared to single-beam devices for all beam lengths. This confirms that the tuning fork structure helps to decouple motion of the beams from the clamps, which effectively increases the clamp stiffness.

\begin{figure}[t]
\centerline{\includegraphics[width=\linewidth]{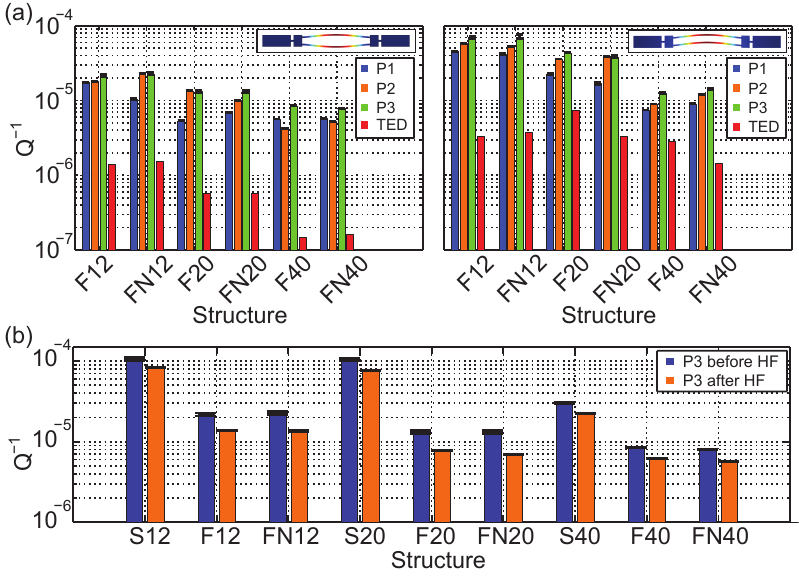}}
\caption{. (a) Experimentally measured $Q_{M}^{-1}$ and FEM calculated $Q_{TED}^{-1}$ for out-of-phase (left) and in-phase (right) mechanical modes. The FEM calculated mode shapes are also shown in the insets. (b) Experimentally measured $Q_{M}^{-1}$ before and after HF treatment. The 95~$\%$ confidence interval ranges for $f_{M}$ and $Q_{M}$, determined by a nonlinear least squares fit to the data, are denoted by the error bars in the graphs. Some of the error bars are smaller than the bar graph line widths.}
\label{fig:fig4}
\end{figure}

According to Fig.~\ref{fig:fig3}(b), the tuning fork design helps to increase $Q_{M}$ for all the beam lengths in the experiment. We believe the reduction of clamping losses is one of the primary reasons for the experimentally measured $Q_{M}$ improvements, based on the following discussion. First, we plot measured $Q_{M}$ values for both the out-of-phase and in-phase mechanical modes for the tuning fork structure in Fig.~\ref{fig:fig4}(a). In all the cases, the out-of-phase modes have higher $Q_{M}$ values than the corresponding in-phase modes, which is expected because the in-phase modes do not produce destructive elastic wave interference in the clamping regions. Next, we note that the tuning fork design is more effective in achieving increased $Q_{M}$ in devices with shorter beams. This is because the fraction of the modal mechanical energy within the clamping regions is comparatively smaller for longer beams, so that for long enough beams, clamping loss is likely no longer a dominant loss mechanism~\cite{ref:Hao_Ayazi}. We also fabricated devices with 80~$\mu$m long beams (data not shown), with measured $Q_{M}$ comparable to the 40~$\mu$m devices. We note that very little difference is seen in the experiments between the $f_{M}$ and $Q_{M}$ values of the tuning forks with and without neck, indicating that the energy dissipation reduction is essentially the same for both.

In addition to the clamping losses, we also consider whether the TED contribution to $Q_{M}$ may be improved by the tuning fork geometries, using FEM simulations in which material deformation is coupled to temperature gradients~\cite{ref:tuning_fork_note}. Indeed, we do find that the TED contribution in tuning fork devices is smaller than that in single-beam ones for all beam lengths, as shown in Fig.~\ref{fig:fig3}(b). Simulation results of the local temperature gradient are shown in Fig.~\ref{fig:fig3}(c). The localized deformation in the tuning fork structure helps to decrease the deformation-induced heating and cooling areas. Therefore, the heat flows driven by the temperature gradient become smaller in the tuning fork structure, which results in lower TED than that of the single-beam structure. In addition, calculation of TED for both the out-of-phase and in-phase modes in tuning fork geometry (Fig.~\ref{fig:fig4}(a)) confirms that the out-of-phase modes have lower TED. However, the TED improvements in simulation are much smaller than the total loss improvements observed in the experiment, which indicates that the $Q_{M}$ improvements are not mainly due to decreased TED.

To simulate the effect of material losses, similar to a previously published work~\cite{ref:unterreithmeier2010damping}, we introduce an additional imaginary part to the Si$_{3}$N$_{4}$ Young’s modulus, so that it becomes $E = E_{1}+iE_{2}$, and determine the resulting quality factor $Q_{mat}$ from finite element simulations. Under the assumption that ($1/Q_{clamp} + 1/Q_{other}$) is small in the tuning fork geometries, we choose a value of $E_{2}~=~20$~MPa ($E_{1}~=~290$~GPa) to achieve a reasonable agreement between experimental losses and simulated ($1/Q_{TED}+1/Q_{mat}$) for the tuning fork devices. This single choice of $E_{2}$ is consistent with the literature~\cite{ref:schmid2011damping} and results in losses that do not exceed the experimentally measured total losses for any of the devices. The simulation results show that the tuning fork structure helps to decrease the damping due to the material defect losses ($1/Q_{mat}$) in addition to decreasing the clamping losses. This can be explained by the simulated elastic energy density shown in Fig.~\ref{fig:fig3}(d), corresponding to the strain distribution during vibration. The deformation of the tuning fork structures is more localized in the center beam region, while the deformation in single beam structures is distributed in both beams and clamps. With a smaller amount of deformed material, damping due to material loss is reduced in tuning fork structures.

As shown in Fig.~\ref{fig:fig3}(b), differences between experimentally measured $1/Q_{M}$ and simulated ($1/Q_{TED}+1/Q_{mat}$) of single beam devices are much larger than those of the tuning fork devices, especially for shorter length beams, indicating higher clamping losses for single beam structures. In total, we believe the clamping loss, damping due to material loss, and TED are all reduced in the tuning fork structure, through localized deformation resulting from elastic wave interference.

From the characterization we also noticed that although they are nominally the same structures, samples from chip P3 have lower $Q_{M}$ than those from other batches. We briefly consider whether this is due to surface losses, which are parts of the material losses. Such losses are caused by the friction processes resulting from surface defects and roughness~\cite{ref:imboden2014dissipation,ref:seoanez2008surface} and have been viewed as a potentially universal limiting loss mechanism in Si$_3$N$_4$ nanomechanical resonators~\cite{ref:villanueva_schmid_surface_loss}, might be reduced in our structures via surface treatment. We immersed the P3 devices into 1:10 hydrofluoric acid: water (HF:H$_2$O) for 90~s to remove a thin layer of material that may have been damaged during the dry etching process. According to Fig.~\ref{fig:fig4}(b), $Q_{M}$ for all of the devices on this sample were improved by up to 2 times after the HF treatment, approaching (but not exceeding) the highest values we have achieved in untreated samples (shown in Fig.~\ref{fig:fig2}). Further investigation is needed to determine whether a surface treatment can be applied to yield higher $Q_{M}$ values than the best untreated samples.

In conclusion, we have developed silicon nitride tuning fork cavity optomechanical transducers. These structures simultaneously increase the resonant frequency ($f_{M}$) and mechanical quality factor ($Q_{M}$) of the fundamental mechanical mode compared with single beam devices, by up to 1.4 times and 12 times, respectively, through an effective increase in the clamp stiffness and reduction in energy dissipation. By engineering the clamp geometry further, calculation indicates that we increase the tensile stress in the beams by 2.9 times compared with the intrinsic stress of the Si$_3$N$_4$ film, resulting in an experimentally measured frequency increase of 1.5 compared with that of the regular tuning fork, while the mechanical damping remained unchanged. The highest measured $f_{M}Q_{M}$ product of $6.35{\times}10^{12}$~Hz at room temperature is on par with the highest values reported for doubly-clamped Si$_3$N$_4$ beams. This tuning fork design with both high $Q_{M}$ and high $f_{M}$ can find applications in force sensing applications, including those that use active feedback cooling to damp mechanical motion~\cite{ref:wilson_kippenberg_feedback_cooling,ref:krause_painter_feedback_cooling}.

The authors acknowledge Dr. Alexander Grey Krause from Delft University of Technology in the Netherlands for helpful discussion.


%

\newpage
\large
\textbf{Supplemental Material}
\normalsize
\setcounter{figure}{0}
\setcounter{equation}{0}
\makeatletter
\renewcommand{\theequation}{S\@arabic\c@equation}
\renewcommand{\thefigure}{S\@arabic\c@figure}

\section{Fabrication}

The devices were fabricated (Figure S1) on commercially available Si$_3$N$_4$ coated silicon wafers. The stoichiometric Si$_3$N$_4$ layer was 250~nm thick and grown via low pressure chemical vapor deposition (LPCVD), with an internal tensile stress of $\approx1.1$~GPa measured by the wafer bowing method. A layer of positive tone electron-beam (E-beam) resist was spin-coated on the Si$_3$N$_4$ film, followed by E-beam lithography at 100~keV and development in hexyl acetate at 8$^{\circ}$C. The patterns were transferred to the Si$_3$N$_4$ film by a CHF$_3$/Ar/O$_2$ inductively-coupled plasma reactive ion etch (RIE). Note that this RIE step left a thin layer of SiO$_2$ on top of the exposed Si, preventing a successful KOH undercut. Thus, we carried out an additional SF$_6$/C$_4$F$_8$ inductively-coupled plasma RIE to remove the SiO$_2$ layer. The resist was then removed by Nanostrip. The fabrication process is finished by device undercut in a 20~$\%$ KOH bath at room temperature and a following N$_2$ blow-dry.

\begin{figure}[h]
\centerline{\includegraphics[width=\linewidth]{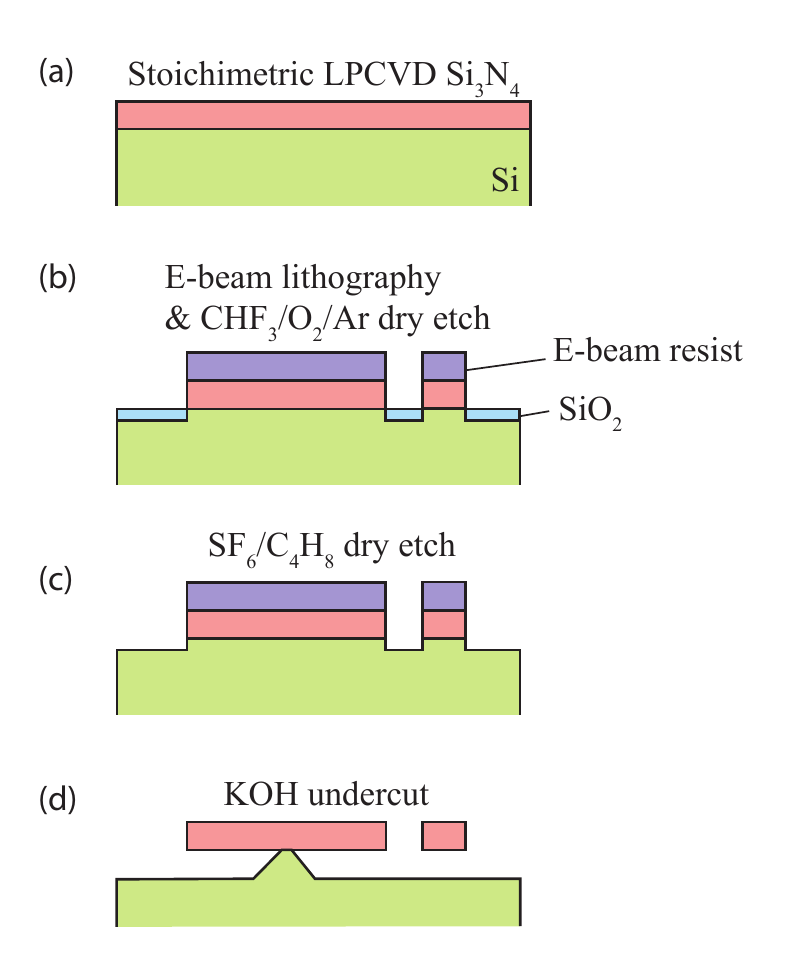}}
\caption{Fabrication Process}
\label{fig:figS1}
\end{figure}

\section{Characterization setup}

Measurements of optical and mechanical modes were performed using the characterization system shown in Figure S2. Light from a 1425~nm to 1470~nm tunable laser was sent through a polarization controller and into an optical fiber taper waveguide~\cite{ref:Srinivasan7,ref:michael_dimple}, which provided evanescent coupling into and out of the microdisk resonator. The device under test was housed in a vacuum chamber at a pressure $\approx0.13$~Pa ($\approx10^{-3}$~torr). Motion of the mechanical resonator introduced a time-varying effective refractive index of the microdisk and hence modulated the optical mode wavelength. Thus, fixing the laser wavelength onto the shoulder of the optical mode resulted in intensity-modulated light transmission from the optical resonator. This output light was split by a fiber coupler, where 10~$\%$ of the power was used for swept-wavelength spectroscopy of the optical cavity modes, while the remaining 90~$\%$ was sent to a photodetector whose electrical output signal was spectrally analyzed. Each mechanical mode was manifested in a strong peak in the radio frequency (RF) spectrum.

\begin{figure}[h]
\centerline{\includegraphics[width=\linewidth]{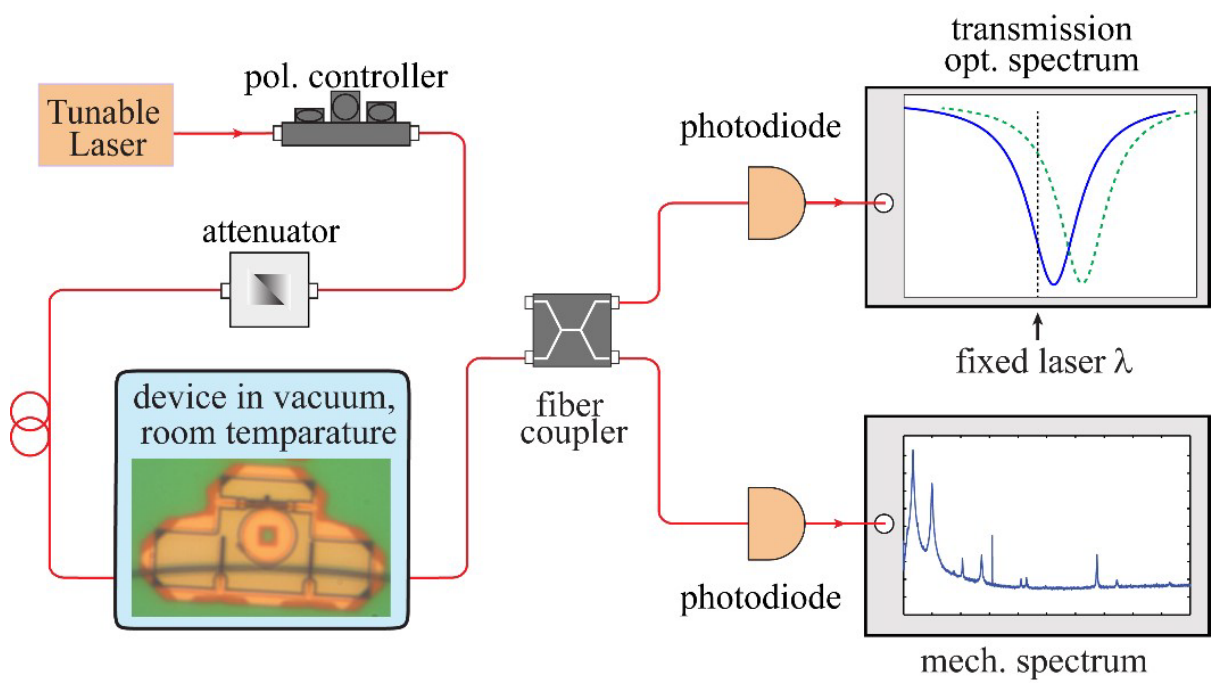}}
\caption{Characterization system}
\label{fig:figS2}
\end{figure}

In this work, we used the fiber taper probe due to the flexibility it provided in enabling measurement of devices across multiple chips. The fiber taper was fabricated by heating and stretching an optical fiber to thin its diameter down to approximately 1 µm. In addition, a local indentation (‘dimple’) with 10 µm radius of curvature was formed at the thinnest region of the taper so that it could be used for selective probing2. In future work focused on force sensing applications, it will be replaced by an integrated on-chip wave guide with pigtailed optical fibers.

\section{Finite element method (FEM) simulation}

In order to better understand the performance of the tuning fork design as well as the experimentally measured spectrum data, we calculated the residual stress, mechanical modes, thermoelastic dissipation (TED), optical modes, and the resulting optomechanical coupling parameters in our devices in a commercial FEM software. The principles of the calculation, as well as the key results of stress tuning, are described in detail below.

\subsection{Stress tuning by the clamp design}

All the devices are doubly clamped to maintain the intrinsic high tensile stress in the LPCVD Si$_3$N$_4$ film (1078 MPa in our samples). The uniformly distributed stress in the film and in the device before undercut will be redistributed after undercut. As a result, the free-standing clamps will shrink after the devices undercut, resulting in the thin beams being stretched, and hence the residual stress is increased (as shown in Figure S3). The residual stress was calculated by a structural mechanics stationary analysis, given a fixed boundary condition at the end of each clamp and the pre-stress in the Si3N4 film (1078 MPa) as the initial condition. The result shows that for all the devices, the tensile stress in the beams became larger while that in the clamps became smaller.

Figure S3 shows the result of the tuning fork and single beam device with 20~$\mu$m beam length. For the single beam device, the stress is increased to 1316 MPa in the beam after undercut, while decreased to 197 MPa in the clamps. Similarly, the tuning fork has a stress of 1318 MPa in the beam and 196 MPa in the clamps. The exaggerated deformation shows the shrinkage of the clamps and the elongation of the beams. A further increase in the beam stress (and corresponding increase in mechanical frequency) can be achieved by modifying the clamp design. To that end, we also designed a high-stress fork which has a 100~$\mu$m long clamp on one side, as shown in Fig.~1(f) in the manuscript. In this device, the beam stress after undercut is calculated to be 3080 MPa, which is approximately 3 times the intrinsic Si$_3$N$_4$ film stress. This stress increase is manifested by an experimentally measured $f_{M}$ increase of 1.5 times and $Q_{M}$ increase of 1.1 times, respectively, compared with regular tuning forks (as shown in Fig.~2 in the manuscript; we note that the regular tuning forks themselves have the aforementioned 1.3 times increase in stress compared to the deposited film). These results confirm that this on-chip, design-enabled stress tuning method can be used to widely adjust the internal tensile stress and the resonant frequency of a nanomechanical resonator, which is of general importance for most of nanomechanical resonators to achieve high $f_{M}$ and $Q_{M}$.  Finally, we note that these results are not specific to the tuning fork geometry, and that this stress tuning method is applicable to any doubly-clamped mechanical resonator with pre-distributed stress.

\begin{figure}[h]
\centerline{\includegraphics[width=\linewidth]{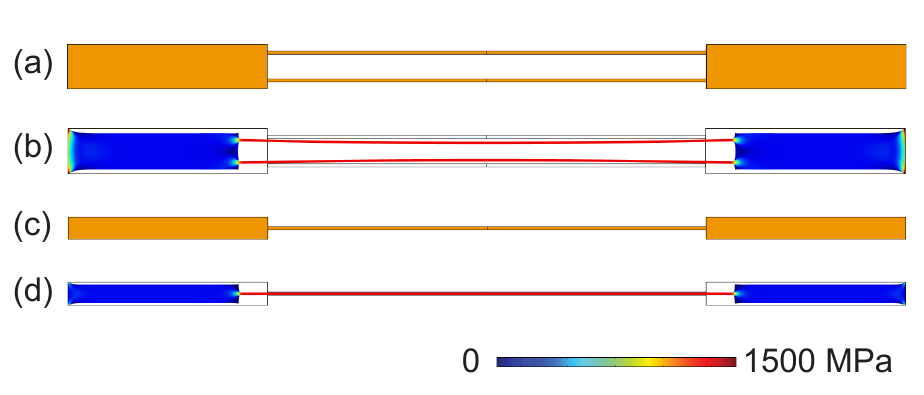}}
\caption{Calculated deformation and corresponding stress re-distribution for the tuning fork structure (a) before and (b) after undercut and those for the single beam (c) before and (d) after undercut. The initial beam length is 20~$mu$m for all the devices shown. The color map shows the stress distribution. The deformation is exaggerated by 70 times.}
\label{fig:figS3}
\end{figure}

\subsection{Calculation of mechanical resonances}

The mechanical modes of the single beam and tuning fork resonators were calculated by solving the following eigenvalue problem for the equation of motion of the displacement field, with the residual tensile stress of the Si3N¬4 film and geometric nonlinearity taken into account:

\begin{equation}
\rho_{0}\frac{\partial^{2} \bf{u}}{\partial t^{2}}-\nabla\cdot\bf{\sigma} = \bf{F}_{v}
\end{equation}

Here, $\rho_{0}$ is the initial material density, $\bf{u}$ is the displacement field, $\bf{\sigma}$ is the elastic part of second Piola-Kirchhoff stress tensor, and $\bf{F}_{v}$ is the volume force vector. A zero displacement boundary condition was applied to the outer ends of the fork clamps.

\subsection{Calculation of TED of mechanical modes}

The corresponding thermoelastic quality factor QTED for each mechanical mode was evaluated by the equation~\cite{ref:Comsol_TED_note}

\begin{equation}
\rho_{0}c_{p}\frac{dT}{dt}=\nabla\cdot(\kappa\nabla T)+Q-T_{0}\biggl(\frac{\partial \epsilon}{\partial T}\biggr)_{\sigma}:\frac{d\sigma}{dt}
\end{equation}

Here, $c_p$ denotes the heat capacity of the solid at constant stress, $T$ the absolute temperature, $\kappa$ tje thermal conductivity, $Q$ the heat source per unit volume, $T_0$ the reference temperature at which the strain and stresses take the initial values, and $\epsilon$ the material stain tensor. Note the last term is a double dot product of two second-order tensors defined as $T:U=\sum_{i}\sum_{j}t_{ij}u_{ij}$, the result of which should be a scalar. The boundary condition we applied to the outer ends of the fork clamps is zero temperature deviation (isothermal boundary condition).

The resonant frequencies of both the in-phase and out-of-phase in-plane modes show general agreement (to within 15~$\%$) between simulation and experiment. The remaining discrepancy may be due to a number of factors, including the fabricated beam widths not matching the design values used in simulation, and imprecise modeling of the clamp region and underlying silicon supports. The latter is particularly important for short single beam devices, for which the specifics of the clamp geometry has a significant influence of mechanical mode frequency. The calculated $Q_{TED}$ values are much larger than the experimentally-measured $Q_{M}$, indicating that thermoelastic dissipation is not the dominant factor contributing to the observed energy decay rates.

\subsection{Calculation of optical resonances}

The whispering-gallery optical modes (WGMs) in the optical resonator can be described by the electromagnetic wave equation

\begin{equation}
\nabla\times(\nabla\times\bf{E})=\epsilon(\omega/c)^2\bf{E}
\end{equation}

Here, $\textbf{E}$ denotes the electric field, $\epsilon$ is the permittivity tensor, $\omega$ is the angular frequency, and $c$ is the speed of light.

Given a specific geometry, the electric field can be obtained by solving the eigenvalue problem in eq. (S1) around a specific wavelength. A perfectly-matched layer was used to simulate open boundaries. In addition, since most of the electric field is concentrated in the periphery of the resonator, the influence from the Si pedestal that supports the microdisk can be ignored in the calculation. For a disk resonator with a diameter $D\approx15$~$\mu$m and Si$_3$N$_4$ thickness $t\approx250$~nm, we found the TE$_{1,43}$ WGM at 1439.89~nm, which has a good match with the experimental result (1441.79~nm).

\subsection{Calculation of the optomechanical coupling parameter}

The optomechanical coupling parameter $g_{OM}$ quantifies the shift in the optical resonance frequency $\omega_{o}$ per unit displacement $x$ of the mechanical resonator, that is, $g_{OM} = \partial\omega_{o}/\partial x$. As presented in Ref.~\cite{ref:eichenfield2}, a perturbation theory for shifting material boundaries~\cite{ref:Johnson6} can be employed to derive the following expression for $g_{OM}$, which can be calculated from the mechanical and optical modes as determined above:

\begin{equation}
g_{OM}=-\frac{\omega_{o}\int dA(\bf{Q}\cdot\bf{n})\biggl(\Delta\epsilon|\bf{E}_{\parallel}|^2-\Delta(\epsilon^{-1})|\bf{D}_{\perp}|^2\biggr)}{2\int dV \epsilon|\bf{E}|^2}
\end{equation}

Here, $\bf{Q}$ is the normalized mechanical displacement field; $\bf{n}$ is the surface normal; $\bf{E}$ and $\bf{D}$ denote the modal electric and electric displacement fields, respectively, $\Delta=\epsilon_{dielectric}-\epsilon_{air}$, and $\Delta(\epsilon^{-1})=\epsilon_{dielectric}^{-1}-\epsilon_{air}^{-1}$, in which $\epsilon_{dielectric}$ and $\epsilon_{air}$ are the permittivities of the tuning fork and air, respectively. To calculate $g_{OM}$, the optical and mechanical modes are separately calculated, yielding $\bf{E}$, $\bf{D}$, and $\bf{Q}$, and the integrals in eq. S4 are evaluated, with $\bf{n}$ obtained from the surfaces of the undeformed structure. The calculated $g_{OM}/2\pi$ is 140~MHz/nm for coupling between the first radial order, transverse electric (TE) polarized disk mode in the 1450~nm band and the fundamental, in-plane and out-of-phase mechanical mode of the 12~$\mu$m long tuning fork geometry ($f_{M}\approx33$~MHz). The calculated $g_{OM}$ values are only slightly smaller ($g_{OM}/2\pi\approx130$~MHz/nm) for the longer 20~$\mu$m long tuning fork geometry ($f_{M}\approx17$~MHz).

We also compare these result with the calculation of $g_{OM}$ by a second approach, in which a deformable mesh is used to simulate the change in optical mode frequency with respect to mechanical displacement. Similar to our previous work~\cite{ref:Liu_Srinivasan_wide_stiffness}, we first simulated the relevant mechanical mode of the tuning fork (in-plane and out-of-phase) to obtain its frequency and mode shape. Next, the tuning fork was deformed with the mechanical mode shape with a prescribed boundary displacement value. The optical modes in the disk with the deformed mechanical resonator were then calculated by solving the electromagnetic eigenvalue problem. We then adjusted the boundary displacement value and re-solve the electromagnetic eigenvalue problem. Repeating this procedure enabled $g_{OM}$ to be obtained as the slope of the fitted frequency-modal displacement curve. The calculated $g_{OM}/2\pi$ for the 12~$\mu$m long tuning fork using this approach is 146~MHz/nm, which has a good match with that calculated by the perturbation theory approach.



\end{document}